\documentclass{IOS-Book-Article}

\usepackage{mathptmx}
\usepackage{multirow}
\usepackage[center]{caption}
\usepackage{graphicx}
\usepackage{url}
\usepackage{cite}
\usepackage{amsmath}
\usepackage{color}
\usepackage{multirow}
\usepackage[bottom]{footmisc}

\def\hb{\hbox to 10.7 cm{}}

\begin{document}
 
\pagestyle{headings}
\def\thepage{}

\begin{frontmatter}              

\title{Methods for Computing Legal Document Similarity: A Comparative Study}


\author[A]{Paheli Bhattacharya}
\author[B]{Kripabandhu Ghosh}
\author[C]{Arindam Pal}
and
\author[A]{Saptarshi Ghosh}

\address[A]{Indian Institute of Technology Kharagpur, India}
\address[B]{Tata Research Development and Design Centre (TRDDC) Pune, India}
\address[C]{Data61, CSIRO and Cyber Security CRC, Sydney, New South Wales, Australia}


\begin{abstract}
Computing similarity between two legal documents is an important and challenging task in the domain of Legal Information Retrieval. Finding similar legal documents has many applications in downstream tasks, including prior-case retrieval, recommendation of legal articles, and so on. 
Prior works have proposed two broad ways of measuring similarity between legal documents -- analysing the precedent citation network, and measuring similarity based on textual content similarity measures. But there has not been a comprehensive comparison of these existing methods, on a common platform. In this paper, we perform the first systematic analysis of the existing methods. In addition, we explore two promising new similarity computation methods -- one text-based and the other based on network embeddings -- which have not been considered till now. 
\end{abstract}

\end{frontmatter}


\section{Introduction}
\label{sec:intro}
\vspace{-3mm}

In countries following the Common Law system, there are two primary sources of law -- Statutes (established laws) and Precedents (prior cases). When a case comes to a legal expert, he has to go through a huge number of legal documents to understand and analyse which ones are relevant to the current case. He prepares his legal reasoning citing or referring these relevant or similar cases. 
But with the advancement of the Web and huge amount of legal content being made available everyday, it is now becoming intractable for legal practitioners to find these relevant legal documents. This calls for the need of automating the search of similar legal documents. 
Automatically retrieving similar documents will also help legal academicians who wish to know about an area of law. 

In this paper, we focus on the task of computing similarity between two legal documents. The notion of similarity is domain-specific and not completely defined; ultimately, two legal case documents are considered similar if legal experts judge them to be similar. The challenge is to automate this similarity computation.

Existing automatic methodologies for finding similar legal documents can be broadly classified into two categories --  (i)~network-based methods (e.g.,~\cite{kumar2011similarity}), which rely on citations to prior case documents, and  (ii)~text-based methods (e.g.,~\cite{contract-semantic-text-matching,mandal2017measuring}, which use the content/textual information of the documents. We refer to these two types of similarity measures as {\it Precedent Citation Similarity} and {\it Textual Similarity} respectively. 
But instead of evaluating the methods on a common dataset of legal documents, different prior works have developed their own set of documents. This situation poses a difficulty in understanding which method is more efficient in finding legal document similarity and why.

In this paper, we aim to bridge this gap. Specifically, we reproduce several existing Precedent Citation Similarity and Textual Similarity methods on a common dataset introduced in~\cite{kumar2013similarity} (which was also used in our prior work~\cite{mandal2017measuring}). 
The dataset contains $47$ pairs of Indian Supreme Court case documents, where similarity between each pair of documents is annotated on a scale from $0-10$ by law experts. We compare all the existing methods of finding legal document similarity on this dataset to ensure a fair comparison.

In addition, we propose two new methodologies for deriving legal document similarity -- 
(i)~a Precedent Citation Similarity-based method using a recent graph embedding approach (Node2Vec~\cite{node2vec}) on the citation network, and 
(ii)~a Textual Similarity method which finds the textual similarity between the different thematic segments (facts, arguments, ratio, ruling etc.) of case documents. 
We also combine various Precedent Citation Similarities and Textual Similarity approaches and analyse their performance.

To the best of our knowledge, this is the first work that (i)~attempts to perform a fair comparison of existing methods for computing legal document similarity, and 
(ii)~introduces additional notions of similarity -- one using a network/graph embedding based approach and another using similarities between the different thematic segments in a document.

The rest of the paper is organized as follows. In Section \ref{sec:dataset}, we describe how we prepare the training and test datasets. In Section \ref{sec:methods}, we explain the existing algorithms for computing similarity between legal documents. In Section \ref{sec:results}, we report and analyse the results obtained from our experiments.  In Section \ref{sec:conclusion}, we summarize the lessons learned in the paper and give some future research directions.
\section{Dataset}
\label{sec:dataset}
\vspace{-3mm}

In this paper, we consider legal judgments from the Supreme Court of India, crawled from the website of Thomson Reuters Westlaw India (\url{http://www.westlawindia.com}). We crawled $53,210$ case documents in total.
Note that we use only the publicly available full text of the judgment. All other proprietary information had been removed before performing the experiments described in this paper.

\vspace{3mm} \noindent 
\underline{\bf Constructing a citation network:}
We construct the prior-case citation network of this document set to compute the Precedent Citation Similarity. The vertices of the network are the case documents. A directed edge exists between two vertices $i$ and $j$ if document $i$ cites document $j$ in its text. 
Consider an example graph shown in Figure~\ref{fig:example}. 
In our example, an edge exists from vertex $A$ to $E$ since $A$ cites $E$. 
There were 53,210 nodes and 208,921 edges in total in this citation graph. 

\vspace{3mm} \noindent 
\underline{\bf Data for Evaluation:} For evaluation of various legal document similarity methods, we consider the dataset introduced in~\cite{kumar2013similarity} (which was also used in our prior work~\cite{mandal2017measuring}). It contains $47$ pairs of Indian Supreme Court case documents. Each pair is assigned a score on a scale of $0-10$ where $0$ represents that the document pair is not similar and $10$ represents that the pair has maximum similarity.
To ensure a fair comparison, we use the same dataset for comparing all the methods explored in this paper.
\vspace{-3mm}
\section{Methods for Legal Document Similarity}
\label{sec:methods}
\vspace{-3mm}

In this section we describe the basic ideas behind the existing methodologies we reproduce in this work.
As stated earlier, the existing methodologies can be broadly classified into two categories --
(i)~methods for precedent citation similarity, and (ii)~methods for textual similarity.

\subsection{Methods for Precedent Citation Similarity}

\begin{figure}[tb]
\centering
\caption{A toy example representing a precedent citation network.}
    \includegraphics[width=6cm,height=3cm]{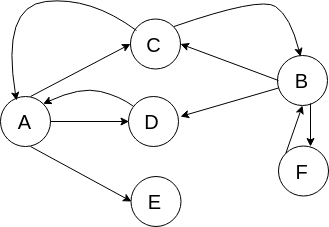}
  \label{fig:example}
\end{figure}

We explain the Precedent Citation Similarity metrics using the running example in Figure~\ref{fig:example}, where we want to measure the similarity between vertices A and B. Apart from three existing network-based similarity measures, we also describe an additional Precedent Citation Similarity method which we newly explore in this work (Node2Vec).

\begin{itemize}
    \item \textbf{Bibliographic Coupling~\cite{kumar2011similarity}} : It is defined as the number of common precedent cases cited by a document pair, i.e, number of common out-citations. In the example graph, the set of out-citations of $A$ is $A_{out} = \left \{ C,D,E \right \}$ and the set of out-citations of $B$ is $B_{out} = \left \{ C,D,F \right \}$. The common out-citation is $A_{out} \cap B_{out} = \left \{ C,D \right \}$. So, the bibliographic coupling based similarity $sim\left ( A,B \right ) = |A_{out} \cap B_{out}| = 2$. This value is normalized by the total number of distinct out-citations of $A$ and $B$, which in this case is $|A_{out} \cup B_{out}| = 4$.
    
    \item \textbf{Co-citation~\cite{kumar2011similarity}} : This metric is defined as the number of common incoming citations to each document of the document pair. In the example graph, the in-citations to $A$, come from the set $A_{in} = \left \{ C,D \right \}$ and in-citations of $B$ come from $B_{in} = \left \{ C,F \right \}$. The common in-citation is $A_{in} \cap B_{in} = \left \{ C \right \}$. So, the co-citation based similarity $sim\left ( A,B \right ) = |A_{in} \cap B_{in}| = 1$. This value is normalized by the total number of distinct in-citations to $A$ and $B$, which in this case is $|A_{in} \cup B_{in}| = 3$.
    
    \item \textbf{Dispersion~\cite{minocha2015finding}} : Dispersion was originally used to identify romantic relationships in the Facebook network. In the context of legal document similarity, it aims to find to what extent the neighbours (out-citation documents) of two documents are themselves similar (occurs in the same community/cluster). We use the \textit{NetworkX} implementation for this measure.\footnote{\url{https://networkx.github.io/documentation/networkx-1.9/reference/generated/networkx.algorithms.centrality.dispersion.html}}
    
    \item \textbf{Node2Vec}: We explore a novel approach based on graph embeddings on the same prior-case citation network on which the above metrics have been implemented. Node2Vec~\cite{node2vec} is a state-of-the-art algorithm for learning embeddings of nodes in a homogeneous network (a network having same type of nodes, e.g. a network of documents citing each other, friendship network of people etc.). Given such a network, Node2Vec aims to map the vertices/nodes of the graph to a vector space such that nodes having similar neighbourhoods in the network have similar embeddings/representations. Given the vector representation of the vertices, we can compute cosine similarity between the vectors to understand the similarity between the two vertices. For our running example, if the vector of vertex $A$ as given by Node2Vec is represented as $\vec{A}$ and vector of vertex $B$ is $\vec{B}$, then $sim\left ( A,B \right ) = \frac{\vec{A}\cdot\vec{B}}{\left \|\vec{A} \right \| \left \| \vec{B} \right \|}.$
\end{itemize}

\subsection{Methods for Textual Similarity}

Now we describe methods for computing textual similarity between legal documents. Along with reproducing methods from prior works, we also propose a new method for textual similarity, namely the similarity between the thematic segments of the case documents.

\begin{itemize}

    \item \textbf{Paragraph Links~\cite{kumar2013similarity}} : In this measure of similarity, a network is formed in which the nodes/vertices are the paragraphs of the documents. Links/edges are established between two vertices/paragraphs, if the similarity between the paragraphs (as measured by TF-IDF) is above a particular threshold. To measure the similarity of the two documents, bibliographic coupling on the above network is calculated. 
    
    \item \textbf{FullText Similarity~\cite{mandal2017measuring}}: Similar to Node2Vec, Doc2Vec~\cite{doc2vec} represents a whole document in a vector space. The vectors preserve the semantics of the document, such that semantically similar documents have similar vector representations. In our prior work~\cite{mandal2017measuring} we explored a wide range of legal document representation methods (e.g., whole document, a summary of the document, reason for citation of prior-judgments) and also state-of-the-art techniques (e.g., TF-IDF, topic models, word embeddings and document embeddings) to calculate  document similarity between various representations. We observed that document embedding using Doc2Vec on the whole document gave the best result in computing legal document similarity. In this paper, we utilise this technique developed in~\cite{mandal2017measuring}. 
    We train Doc2Vec on a large set of Indian Supreme Court case documents (the 53,201 documents stated in Section~\ref{sec:dataset}, excluding the documents in the dataset for evaluation). We then use the learned model to infer the vectors of the documents of the $47$ pairs in the evaluation dataset. We then compute cosine similarity between the vectors of the documents to find the similarity.
    
    \item \textbf{Thematic Similarity}: It is known that a legal case document contains various themes/segments/functional parts like Facts, Arguments, Ratio of the decision, Final judgment and so on. While judgments of many countries reflect this segmented structure through section headings, Indian legal judgments are devoid of any such systematic structure. 
    We recently proposed a Machine Learning-based method for thematic segmentation of Indian Supreme Court Case documents~\cite{bhattacharya2019identification}. 
    The method (based on deep neural networks) has been shown to work well for 7 rhetorical roles/themes (Facts, Arguments, Ratio of the decision, Statute, Precedent, Ruling by Lower Court, and Ruling by Present Court) over documents from 5 popular legal domains. 
    The implementation of this segmentation method is publicly available at \url{https://github.com/Law-AI/semantic-segmentation}.
    
    We use the same method to segment the two documents for which we want to calculate the similarity. After getting the segments (Fact, Argument, Ratio, etc.) from both the documents, segment-level similarities (e.g., Fact-Fact similarity, Argument-Argument similarity, Ratio-Ratio similarity, and so on) between the two documents are computed. Then an aggregated similarity is reported as the final similarity between the document pairs. 
    For aggregation we try the following two ways -- 
    (i)~\textit{max}: the maximum similarity value between any segment of the document pairs is considered as the overall similarity, and 
    (ii)~\textit{average}: the average similarity across the segments is considered as the overall similarity of the document pair.
\end{itemize}

\noindent In this section, we have described several Precedent Citation Similarity measures and several Textual Similarity measures for computing the similarity between two legal documents. We will compare these methodologies (and also their combinations) in the next section.
\vspace{-3mm}
\section{Results and Analysis}
\label{sec:results}
\vspace{-3mm}

We now compare the performances of the various methods stated in the previous section, over the evaluation dataset of 47 document pairs (stated in Section~\ref{sec:dataset}).

\subsection{Method for evaluation}
\vspace{-3mm}
Recall from Section~\ref{sec:dataset} that the evaluation dataset contains 47 document pairs, where each pair has a similarity score in the range $[0,10]$ that is assigned by legal experts. 
To judge the performance of a particular similarity method, 
we use the Pearson Correlation coefficient between the expert scores and the computationally obtained similarities from the said methodology. 

Table~\ref{tab:sample} shows a sample of the evaluation framework. For each document pair, an expert similarity score in the range $0-10$ is available. Each of the methodologies Node2Vec, FullText Similarity, Thematic Similarity, etc assigns a similarity value to the same document pairs. We then compute an overall Pearson Correlation coefficient for each of the methodologies.

\begin{table}[tb]
\caption{A sample of the dataset showing the expert score and the similarity inferred by a subset of methods explored in this paper. The values nearest to the expert score are bold-faced.}

\label{tab:sample}
\begin{tabular}{|c|c|c|c|c|c|}
\hline
\textbf{Document Pair} & \textbf{Expert Score} & \textbf {Node2Vec} & \textbf{\begin{tabular}[c]{@{}c@{}}FullText\\ Similarity~\cite{mandal2017measuring} \end{tabular}} & \textbf{\begin{tabular}[c]{@{}c@{}}Thematic\\ Similarity (Avg)\end{tabular}} & \textbf{\begin{tabular}[c]{@{}c@{}}Thematic\\ Similarity (Max)\end{tabular}} \\ \hline

1992\_47 \& 1992\_76   & 0   & 0.195  & 0.188  & \textbf{0.154}   & 0.571   \\ \hline

1979\_110 \& 1989\_233 & 3   & 0.613   & 0.465  & 0.104    & \textbf{0.415}    \\ \hline

1953\_24 \& 1957\_52   & 7   & 0.234   & 0.264  & 0.377    & \textbf{0.757}  \\ \hline

1983\_37 \& 1979\_33   & 10  & 0.574   & \textbf{0.711}    & 0.209  & 0.692     \\ \hline
\end{tabular}
\end{table}

\begin{table}[tb]
\caption{Pearson Correlation between the Expert Score and Similarity inferred by each method}
\label{tab:results}
\begin{tabular}{|c|c|c|}
\hline
\textbf{Category}                                                                                                    & \textbf{Method}                                                              & \textbf{Correlation} \\ \hline
\multirow{4}{*}{\textbf{\begin{tabular}[c]{@{}c@{}}Prior-case\\ citation\\ network based\\ measures\end{tabular}}} & \begin{tabular}[c]{@{}c@{}}Bibliographic Coupling~\cite{kumar2011similarity} \end{tabular}             & 0.443                \\ \cline{2-3} 
                                                                                                                     & Cocitation~\cite{kumar2011similarity}                                                                       & 0.205                \\ \cline{2-3} 
                                                                                                                     & Dispersion~\cite{minocha2015finding}                                                                    & 0.246               \\ \cline{2-3} 
                                                                                                                     & Node2Vec                                              & \textbf{0.487}                 
                                                                    \\ \hline \hline
\multirow{4}{*}{\textbf{\begin{tabular}[c]{@{}c@{}}Text based\\ measure\end{tabular}}}                               & Paragraph Links~\cite{kumar2013similarity}                                                                                    & 0.33                 \\ \cline{2-3} 
                                                                                                                     & FullText Similarity ~\cite{mandal2017measuring}                                                                   & \textbf{0.605}       \\ \cline{2-3} 
                                                                                                                     & Thematic Similarity (avg) & 0.523                \\ \cline{2-3} 
                                                                                                                     &                                                                     Thematic Similarity (max)              & \textbf{0.599}          \\ \hline
\end{tabular}
\end{table}

\subsection{Analysis}
\vspace{-3mm}

Table~\ref{tab:results} shows the results (Pearson Correlation) for each method. We find that among the Precedent Citation Similarity based methods, the graph embedding approach, Node2Vec, performs the best. Legal citation networks are very sparse. Only a few cases are cited for each document. Results suggest that simpler measures like co-citation and dispersion perform poorly in such scenarios. On the other hand, graph embedding based method could perform much better in comparison.

Among Textual Similarity measures, we find that document embedding on the full text performs the best.\footnote{Note that the prior work~\cite{mandal2017measuring} reported a slightly different Pearson correlation value than what we are reporting here (for the same evaluation dataset), because the training data for the Doc2vec model is different in this work.} 
Thematic similarity combined using the \textit{max} shows very comparable performance. This observation suggests that even textually, there are certain themes in which two documents can be more similar than in others; for instance, two documents may be very similar in terms of their facts but different in their final judgment or arguments. 
It depends on the perspective through which a legal document being similar or dissimilar is judged. In our framework we find that the documents are most similar in Ratio (correlation 0.45), Precedent (0.42) and Fact (0.37). Detailed results are omitted due to space constraints.

\begin{table}[tb]
\caption{Correlation Results after combining Precedent Citation Similarity methods and FullText Similarity}
\label{tab:combo}
\begin{tabular}{|c|c|c|}
\hline
\textbf{Combination}              & \textbf{Methods Combined}                                        & \textbf{Correlation} \\ \hline
\multirow{4}{*}{\textbf{Max}}     & Biblio + FullText Sim                                                       & \textbf{0.626}       \\ \cline{2-3} 
                                  & Cocitation + FullText Sim                                                       & 0.582                \\ \cline{2-3} 
                                  & Dispersion + FullText Sim                                                        & 0.600                  \\ \cline{2-3} 
                                  & Node2Vec + FullText Sim & 0.595                
                                  \\ \hline
\multirow{5}{*}{\textbf{Average}} & Biblio + FullText Sim                                                      & 0.575                \\ \cline{2-3} 
                                  & Cocitation + FullText Sim                                                       & 0.432                \\ \cline{2-3} 
                                  & Dispersion + FullText Sim                                                        & 0.507                \\ \cline{2-3} 
                                  & Node2Vec + FullText Sim & 0.552                \\ \hline
\end{tabular}
\end{table}

\begin{table}[tb]
\caption{Correlation Results after combining Precedent Citation Similarity methods with Thematic Similarity}
\label{tab:combothem}
\begin{tabular}{|c|c|c|}
\hline
\textbf{Combination}              & \textbf{Methods Combined}                                              & \textbf{Correlation} \\ \hline
\multirow{4}{*}{\textbf{Max}}     & Biblio + Thematic Sim                                              & 0.604                \\ \cline{2-3} 
                                  & Cocitation + Thematic Sim                                               & 0.560                 \\ \cline{2-3} 
                                  & Dispersion + Thematic Sim                                                & 0.586                \\ \cline{2-3} 
                                  & Node2Vec + Thematic Sim & 0.530                 \\ \hline
\multirow{4}{*}{\textbf{Average}} & Biblio + Thematic Sim                                              & 0.565                \\ \cline{2-3} 
                                  & Cocitation + Thematic Sim                                               & 0.416                \\ \cline{2-3} 
                                  & Dispersion + ThematicSim                                                & 0.485                \\  \cline{2-3} 
                                  & Node2Vec + Thematic Sim & \textbf{0.615}       \\ \hline

\end{tabular}  
\end{table}

\vspace{3mm} \noindent
\textbf{Combining Textual Similarity and Precedent Citation Similarity:}
We also attempt to combine the Textual and Precedent Citation similarities. We combine FullText Sim (obtained using~\cite{mandal2017measuring}) and Thematic Similarity  -- the two best-performing textual similarity measures -- with the various Precedent Similarity measures. 
The results are shown in Table~\ref{tab:combo} (for combining FullText similarity) and Table~\ref{tab:combothem} (for combining Thematic similarity) respectively. 
We try two aggregation functions namely \textit{max} and \textit{average}. Suppose for a document pair $A$ and $B$, the network based similarity is $s_1$ and text based similarity is $s_2$. Then the aggregated similarity using \textit{max} is $max\left ( s_1,s_2 \right ) $. The aggregated similarity using \textit{average} is $\frac{1}{2}(s_1 + s_2)$.

We find that the \textit{max} performs consistently well in Table \ref{tab:combo}. The best performance is observed when we combine the bibliographic coupling based precedent citation similarity with the FullText similarity. 
On the other hand when we combine Thematic Similarity with the Precedent Citation Similarity measures, the average combination with the graph embedding based approach performs the best.
In both cases, we get a higher correlation with expert scores by combining Textual Similarity and Precedent Citation Similarity, than what we obtained by using one type of methodology alone.

These results show that 
both textual similarity and precedent citation similarity are helpful in measuring the similarity between legal case documents. 
\vspace{-3mm}
\section{Conclusion and Future Work}
\label{sec:conclusion}
\vspace{-3mm}

In this paper, we attempt to compare existing approaches for legal document similarity. We perform a systematic comparison of the methods, on a set of $47$ document pairs. We also explore two new methods, one based on graph embeddings, and the other based on textual similarity between thematic segments. 

We show that understanding legal document similarity is indeed a challenging task, and has various different facets, such as similarity in precedent citation, similarity in textual content, similarity of themes, and so on. Multiple facets of document properties contribute to the overall similarity of the documents.  It may not be possible to capture all of them by a single methodology; rather, an aggregation of the perspectives is necessary.

There are several future directions of the work reported in this paper. 
One immediate future work is to develop better methods for judging similarity of legal case documents, which would agree more closely with the opinion of the legal experts.
A detailed inspection of document-pairs that the experts judge to be highly similar may reveal other factors that need to be considered while judging the similarity of such documents.
In this context, it is important to consider explainability of the similarity computation -- ideally, experts can be asked to explain how they judge similarity between two legal documents, and automatic methods may be developed based on these explanations. We plan to explore these directions in future.

\bibliography{bibliography} 
\bibliographystyle{ieeetr}

\end{document}